\documentclass[apj,twocolumn]{aastex63}

\usepackage{graphicx}  % Include figure files
\usepackage{color}     % for preprints
\usepackage{amsmath}   % for equations. 
\usepackage{enumerate}
\usepackage{lipsum}
\usepackage[makeroom]{cancel}
\usepackage[caption=false]{subfig}

\def\dr{\text{d}r}

\def\STIR/{\textit{STIR}}
\def\vturb{v_{\text{turb}}}
\def\omegaBV{\omega_{\textsc{\scriptsize bv}}}
\def\aMLT{\alpha_{\textsc{\scriptsize mlt}}}

\def\BV/{Brunt-V\"{a}is\"{a}l\"{a}}
\newcommand{\pder}[2]{\frac{\partial#1}{\partial#2}}
\newcommand{\avg}[1]{\langle #1 \rangle}

\submitjournal{ApJ}

\shorttitle{1D Turbulence in GR}
\shortauthors{Boccioli et al.}

\begin{document}
\doublespace

\title{General Relativistic Neutrino-Driven Turbulence in One-Dimensional Core-Collapse Supernovae}

\correspondingauthor{Luca Boccioli}
\email{lbocciol@nd.edu}

\author[0000-0002-4819-310X]{Luca Boccioli}
\affiliation{Center for Astrophysics, Department of Physics, University of Notre Dame, 225 Nieuwland Science Hall, Notre Dame, IN 46556, USA}

\author[0000-0002-3164-9131]{Grant J. Mathews}
\affiliation{Center for Astrophysics, Department of Physics, University of Notre Dame, 225 Nieuwland Science Hall, Notre Dame, IN 46556, USA}

\author[0000-0002-8228-796X]{Evan P. O'Connor}
\affiliation{The Oskar Klein Centre, Department of Astronomy, Stockholm University, AlbaNova, SE-106 91 Stockholm, Sweden}

\begin{abstract}
Convection and turbulence in core-collapse supernovae (CCSNe) are inherently three-dimensional in nature.  However, 3D simulations of CCSNe are computationally demanding.  Thus, it is valuable to modify simulations in spherical symmetry to incorporate 3D effects using some parametric model.  In this paper, we report on the formulation and implementation of general relativistic (GR) neutrino-driven turbulent convection in the spherically symmetric core-collapse supernova code \texttt{GR1D}.  This is based upon the recently proposed method of Supernova Turbulence in Reduced-dimensionality (\STIR/) in Newtonian simulations from \cite{Couch2020_STIR}. When the parameters of this model are calibrated to 3D simulations, we find that our GR formulation of \STIR/ requires larger turbulent eddies to achieve a shock evolution similar to the original \STIR/ model. We also find that general relativity may alter the correspondence between progenitor mass and successful vs.~failed explosions.
\end{abstract}


\section{Introduction} 
\label{sec:intro}
Despite the fact that core-collapse supernovae (CCSNe) have been the focus of computational simulations for more than 50 years,  many fundamental details of the mechanism driving the explosion of  massive stars remain unknown. Nevertheless, our knowledge has significantly improved since the first theoretical \citep{Colgate_White1966,Arnett1966} and observational \citep{Sonneborn1987a} efforts. This has been aided by an exponential growth of computing power.

In the early '80s, one-dimensional (1D) spherically symmetric simulations revealed the crucial role of neutrinos in triggering the explosion through the so-called delayed neutrino-heating mechanism \citep{Bethe_Wilson1985,Bruenn1985}. A few years later, the first two-dimensional (2D) axi-symmetric numerical investigations of CCSNe were also performed \citep{Miller1993_first2D,Herant1994_first2D}, while more recently the extension to three dimensions (3D) has finally become feasible \citep{Fryer2002}. Since then, \hbox{multi-D} simulations have continued to improve, and have shed light on the details of the explosion mechanism (cf. review in Ref.~\cite{Muller2016_review}). For example, the importance of neutrino-driven turbulent convection is now widely recognized \citep{Radice2016,Radice2018_turbulence,Mabanta2018_MLT_turb}. Moreover, new phenomena such as the Standing Accretion Shock Instability (SASI) \citep{Blondin2003}, the Lepton-number Emission Self-sustained Asymmetry (LESA) \citep{Tamborra2014}, Collective Neutrino Oscillations (CNO) \citep{Duan2010} and others, have been  revealed in multi-D simulations. However, the  impact of these phenomena on the explosion is still a topic of active research.

After many years of theoretical and observational work, it is now well established that spherically symmetric 1D simulations  only lead to self-consistent explosions of low mass (typically around 9-11 M$_\odot$) progenitors. These stars  develop an oxygen-neon-magnesium core, whose collapse is triggered by electron captures on Ne and Mg. Contrary to these so-called Electron Capture Supernovae (ECSNe), the more common Fe-core CCSNe do not self-consistently explode in  spherically symmetric simulations that employ a modern treatment of the nuclear equation of state (EOS) and neutrino interactions, with the possible exception of a zero-metallicity, 9.6 M$_\odot$ progenitor \citep{Melson2015_9.6_expl,Betranhandy2020_pair}.
Nevertheless, there have been a number of simulations in  2D and 3D  that have led to successful explosions \citep{Muller2012_2D_GR,Lentz2015_3D,Janka2016_success_expl,Bruenn2016_expl_en,OConnor2018_2D_M1,Muller2019_3Dcoconut,Burrows2020_3DFornax}. However, axial symmetry has recently been shown \citep{Couch2015_turbulence} to favor explosions by artificially enhancing neutrino-heating behind the shock  through an inverse turbulence cascade that is not present in 3D. Therefore, it is generally believed that the final explanation as to what causes the explosion must come from simulations in three spatial dimensions.

Unfortunately, despite the  technological improvements of the last few decades, three-dimensional core-collapse supernova simulations still pose a difficult computational challenge, even for modern supercomputers. In addition to this, 3D simulations from different groups only agree for a few tens of milliseconds after bounce \citep{Cabezon2018_3Dcomparison}. Afterward, they begin to significantly deviate from each other, even when they start from the same initial conditions. The more accessible 2D simulations, that require a relatively affordable CPU time, are also still complicated enough to the point that different groups often find different outcomes for the explosion (see Table 1 from \cite{OConnor2018_2D_M1}), although some promising benchmarking work has already been done \citep{Pan2019_nu_transport_2D_comparison}.

In comparison, modern 1D simulations are much faster to run and more consistent across different codes \citep{OConnor2018_comparison}. Hence, when starting from the same initial conditions, most groups obtain similar results (while older 1D simulations, with outdated sets of neutrino interactions, lead to some inconsistencies [cf. \cite{Mathews2020_nurelic}]). This guarantees a somewhat solid foundation and makes 1D simulations an ideal bench-marking tool for studying how different input physics can affect supernova explosions.

Precisely due to the low computational cost associated with 1D simulations, there have been several attempts to artificially drive explosions in 1D codes that would otherwise yield failed supernovae. The advantage of having models capable of artificially driving explosions is quite straightforward: if the models are \textit{reliable}, one can efficiently explore the parameter space of different physical phenomena. For example, one could explore the connection between the explodability of CCSNe and the progenitor mass \citep{OConnor2011_explodability,Ugliano2012,Pejcha2015_explodability,Ertl2016_explodability,Sukhbold2016_explodability,Ebinger2020_PUSH,Couch2020_STIR}, or the impact of different input physics, such as the nuclear EOS \citep{Schneider2019,Olson2016} and the weak interactions \citep{Fischer2020_nuclear_clusters,Guo2020_muon,Betranhandy2020_pair}) on the explosion itself. Furthermore, one can quantify the dependence of various nucleosynthesis processes on the progenitor mass, or calculate the diffuse relic neutrino background [c.f. review in \cite{Mathews2020_nurelic}].  Moreover, 1D simulations can efficiently study any aspect of supernovae that requires a large sample of progenitors and/or evolution to late times post bounce.

A number of  artificially driven explosion mechanisms have been proposed  over the past few decades, based upon  different parametric models \citep{Blinnikov1993_bomb,Woosley1995_pistons,Ugliano2012,Perego2015_PUSH1,Sukhbold2016_explodability,Couch2020_STIR}. For most of them, the parameters were chosen by comparing the explosion energies and other relevant observables to known measured properties of SN 1987a \citep{Sonneborn1987a} and other SNe. 

The advantage of comparing with observations is that one can directly test the results with respect to real data. On the other hand, since these parametric models are not derived from a physical mechanism, one cannot meaningfully compare these results to 3D simulations and gain insight on the physics of the explosion. Therefore, a different (and complementary) way of approaching the problem is to develop a parametric model based upon a known physical mechanism that is motivated by the success of  3D simulations \citep{Muller2016_Oxburning,Couch2020_STIR}.

An obvious candidate for such a mechanism is neutrino-driven turbulent convection, which plays an important role in CCSNe \citep{Couch2015_turbulence,Radice2016,Radice2018_turbulence,Mabanta2018_MLT_turb}.   Recently,  \cite{Couch2020_STIR} (hereafter CWO20) developed a model called \textit{STIR} (Simulated Turbulence In Reduced-dimensionality), based upon a modified Mixing Length Theory (MLT), that simulates features of 3D turbulence in spherically symmetric simulations.

\textit{STIR} is a Newtonian model for turbulent convection, and the simulations from CWO20 only partially include general relativistic effects through a General Relativistic Effective Potential (GREP) from \cite{Marek2006}. However, we know that General Relativity (GR) plays an important role in the explosion of supernovae \citep{Wilson2003, Muller2012_2D_GR}. Hence, simulations in full GR are desirable. 

In this paper we extend the \textit{STIR} model to a general relativistic treatment.  We then  analyze the differences between full GR and GREP. In Section 2 we describe the essential physics of STIR and its generalization to GR.  Results and a comparison to 3D simulations are described in Section 3, followed by a study of explodability vs.  progenitor mass in Section 4.  Conclusions are given in Section 5. Throughout the manuscript, we chose to adopt natural units, i.e. $G = c = M_\odot = 1$.

\section{Methods}
\subsection{The STIR model of Couch et al. (2020)}
The importance of convection in CCSNe was recognized very early on, and the first attempts at applying MLT to CCSN date back to \cite{Wilson988} and \cite{Bruenn1995}. We follow a more modern  version of those models, that can be found in \cite{Mabanta2018_MLT_turb}, \cite{Mabanta2019_turbulence_in_CCSN} and CWO20. We highlight that this formulation does consider both pressure and composition gradients (see Appendix A of \cite{Muller2016_Oxburning} for an explicit proof of this). However, it doesn't capture lepton number-driven convection, relevant at the surface of the PNS, where neutrinos are trapped below the neutrinosphere and free streaming above it, creating a gradient which could drive convection. However, given the complicated thermodynamics derivatives involved in the modeling of this type of convection \citep{Wilson2003}, we leave its treatment to a future work.

As described in the aforementioned papers, the effects of turbulence can be treated as a perturbation on a background fluid. This can be done by decomposing a generic fluid variable $\psi$ into a mean component $\psi_0$ and a perturbed component $\psi'$: $\psi = \psi_0 + \psi'$, with $\avg{\psi} = \psi_0$, where $\avg{...}$ represents an average over a proper spatial and time domain. Because of the chaotic nature of the motions that characterize turbulence, the average should vanish (i.e.~$\avg{\psi'}=0$). This decomposition, called Reynolds averaging, can be applied to the compressible Euler equations to derive conservation laws that contain explicit contributions from the turbulent motion for total energy, momentum and mass.  In the Newtonian limit, these are:

\begin{gather}
    \label{eq:Reynold1}
    \partial_t \avg{\rho} + \nabla \cdot (\rho_0 \mathbf{v}_0) = 0~~, \\
    \label{eq:Reynold2}
    \begin{split}
    \partial_t \avg{\rho \mathbf{v}} + \nabla \cdot (\rho_0 \mathbf{v}_0 \times \mathbf{v}_0 + P_0 \mathbf{I}) = &-\rho \mathbf{g} \\
    &-\nabla \cdot \avg{\rho \mathbf{R}}
    \end{split} ~~,\\
    \label{eq:Reynold3}
    \begin{split}
    \partial_t \avg{\rho \epsilon} &+ \nabla \cdot \mathbf{u_0} (\rho_0 \epsilon_0 + P_0) \\
    = &-\rho_0\mathbf{v}_0 \cdot \mathbf{g} -\nabla \cdot \mathbf{v}_0 \avg{\rho \mathbf{R}} - \nabla \cdot \mathbf{F}_e + \rho_0 \epsilon'~~.
    \end{split}
\end{gather}
Here, $\rho$ is the mass density, $\mathbf{v}$ is the velocity, $P$ is the pressure, $\mathbf{I}$ is the identity matrix, $\mathbf{g}$ is the gravitational acceleration, $\epsilon$ is the internal energy per unit mass, $\mathbf{R}$ is the Reynolds stress tensor and $\mathbf{F}_e$ is the energy flux due to turbulence. In the above equations, we have neglected terms that are proportional to the turbulent pressure.  This is a good approximation in regimes of low Mach numbers (for more details on the derivation of equations \eqref{eq:Reynold1}-\eqref{eq:Reynold3} see \cite{Murphy2011}, \cite{Mabanta2018_MLT_turb}, and CWO20).

In addition to Eqs. \eqref{eq:Reynold1}-\eqref{eq:Reynold3}, one needs another equation to describe the evolution of the specific turbulent kinetic energy $K=\text{Tr}(\mathbf{R})/2$. The derivation of this equation can be found in \cite{Murphy2011} and CWO20. Here, we simply highlight the important steps that are relevant to our model.

For isotropic turbulence in spherical symmetry, the evolution equation for the turbulent kinetic energy is:

\begin{equation}
    \label{eq:turb_kin_ene}
    \begin{split}
    \pder{\avg{\rho K}}{t} &+ \frac{1}{r^2} \pder{}{r} [r^2 ( \avg{\rho K} v_r + \avg{\rho K v_r'})] \\
    = &- \avg{\rho \mathbf{R}_{rr}} \pder{v_r}{r} + \avg{\rho' v_r'}g - \rho \epsilon'~~.
    \end{split}
\end{equation}

One can rewrite almost all of the unknown expressions in \eqref{eq:turb_kin_ene} in terms of gradients and divergences of $\vturb \equiv v'$ and  $\epsilon_\text{turb} \equiv \epsilon'$, where:

\begin{equation}
    \epsilon_\text{turb} = \frac{v_\text{turb}^3}{\Lambda}
\end{equation}
with $\Lambda$ being the largest turbulent eddy size. The only quantities that remain to be determined are the buoyancy force $\avg{\rho'v_r'}g$ and $\Lambda$. Therefore, to close the system of equations, one needs an expression that relates the transport of turbulent energy to the typical speed of a buoyant convective element. To do this, one refers to MLT, for which the buoyant force that makes the convective element rise through the stratified fluid is:

\begin{equation}
    \avg{\rho' v'} g \approx \rho \vturb \omegaBV^2 \Lambda_{\text{mix}}
\end{equation}
with:

\begin{align}
    \label{eq:lambda_mix}
    \Lambda \equiv \Lambda_{\text{mix}} &= \aMLT \frac{P}{\rho g},\\
    \label{eq:BV_eff}
    \omegaBV^2 &=  g_\text{eff} \left(\frac{1}{\rho} \pder{\rho}{r} - \frac{ 1}{ \rho c_s^2} \pder{P}{r} \right).
\end{align}
In the above equations, $\Lambda_{\text{mix}}$ is the mixing length, $\omegaBV$ is the \BV/ frequency, $c_s$ is the sound speed, and $g_\text{eff}$ is the magnitude of the local effective acceleration. For a fluid in hydrostatic equilibrium, $g_\text{eff}$ simply reduces to the local gravitational acceleration $g$. More in general, however, in the rest frame one should take the acceleration of the fluid into account. Therefore, the total acceleration $g_\text{eff}$ can be expressed as:

\begin{equation}
    \label{eq:geff}
    g_\text{eff} = g - v_r \pder{v_r}{r}~,
\end{equation}
as described in CWO20, where $v_r$ is the radial velocity of the fluid.
During the pre-bounce and early post-bounce phases, both $v_r$ and  $\partial v_r / \partial r$ are small, so that $g_\text{eff}$ simply reduces to $g$. Once the explosion sets in, however, the contribution from the acceleration of the fluid is non-negligible, hence $g_\text{eff}$ becomes smaller than $g$. This effectively shuts off turbulent convection in the exploding material by driving $\omegaBV$ to zero.

The mixing length $\Lambda_\text{mix}$ is the average distance that a convective element will travel before being mixed with (and increasing the internal energy of) the surrounding material. The \BV/ frequency $\omegaBV$ is the rate at which the convective elements are rising. As one can notice from Eq.~\eqref{eq:BV_eff}, $\omegaBV^2$ can be either positive or negative: when $\omegaBV^2>0$ the fluid is convectively unstable, i.e. convection is generated; when $\omegaBV^2<0$ the fluid is convectively stable, i.e. convection is damped. Ultimately, the main parameter of the model is $\aMLT$, which scales the mixing length to the pressure scale height in Eq.~(\ref{eq:lambda_mix}).   Typically $\aMLT \sim O(1)$.

After closing the system of equations derived from the Reynolds decomposition, one obtains the following evolution equation for the turbulent kinetic energy:

\begin{equation}
\begin{split}
\label{eq:vturb}
\pder{\rho \vturb^2}{ t} &+ \frac{1}{r^2}\pder{}{r} [r^2(\rho \vturb^2 v_r - \rho D_K \nabla \vturb^2)] \\ 
&= -\rho \vturb^2 \pder{v_r}{r} + \rho \vturb \omegaBV^2 \Lambda_{\text{mix}} - \rho \frac{\vturb^3}{\Lambda_{\text{mix}}}~~.
\end{split}
\end{equation}
In the above equation, $D_K$ is a diffusion coefficient due to turbulence, defined as:

\begin{equation}
    D_K = \alpha_K \vturb \Lambda_{\text{mix}}~.
\end{equation}
Similar terms appear in the internal energy, electron fraction, and neutrino energy density evolution equations (for the complete set of hydrodynamic equations used in the model, see Eqs. 25-29  and 33 in CWO20). Therefore, strictly speaking, \STIR/ has 4 additional free parameters: $\alpha_{K}$, $\alpha_e$, $\alpha_{\text{Ye}}$, $\alpha_\nu$. However, the convective motions are not very sensitive to the value of these parameters, so we set them to $1/6$ for simplicity, consistent with the choices of \cite{Muller2016_Oxburning} and CWO20.

In the next section we  describe a general relativistic version of the model described above. 

\subsection{STIR in General relativity}
\label{sec:STIR_GR}
The first attempts to create a general relativistic model for convection date back to \cite{Thorne1966}. We will follow the same approach, but using a slightly different formalism. 

All the simulations described in this paper were run with the open-source, spherically symmetric, general relativistic code \texttt{GR1D} \citep{OConnor2010}. The Boltzmann equation for neutrino transport is solved using an M1-scheme, with opacity tables generated using the open-source code \texttt{NuLib} \citep{OConnor2015}.

The metric evolved in \texttt{GR1D} is Schwarzchild-like:

\begin{equation}
    \label{eq:metric}
    \begin{split}
    ds^2 &= g_{\mu\nu} x^\mu x^\nu \\
         &= -\alpha(r,t)^2 dt^2 + X(r,t)^2 dr^2 + r^2 d\Omega^2,
    \end{split}
\end{equation}
where $\alpha$ and $X$ can be expressed as functions of a metric potential $\phi$ (which reduces to the Newtonian potential in the Newtonian limit) and the enclosed gravitational mass $M_\text{grav}$:

\begin{equation}
    \label{eq:alphaX}
    \begin{split}
    \alpha(r,t) &= \exp[ \phi(r,t) ], \\
    X(r,t) &= \left( 1-\frac{2M_\text{grav}(r,t)}{r} \right)^{-1/2}
    \end{split}
\end{equation}

For the present work, we first note that turbulence is mostly relevant far from the proto-neutron star (PNS) where GR effects can be treated as a perturbation. Therefore, one can simply make a few changes to the terms in Eq.~\eqref{eq:vturb} without having to re-derive the entire Reynold's decomposition. The expression for $\omegaBV^2$, however, must be carefully re-derived (a formal derivation of $\omegaBV^2$ for a fluid in hydrostatic equilibrium can be found in Appendix \ref{appendixBV}). Far from the PNS, we invoke the following:
\begin{enumerate}[(i)]
    \item replace the conserved variable $\rho$ with its GR counterpart, i.e. $D=W X \rho$, where \\$W = (1-v^2)^{-1/2}$ and $v = X v_r$;
    \item multiply the RHS of Eq.~(\ref{eq:vturb}) by $\alpha X$. 
    \item multiply the spatial flux in Eq. \eqref{eq:vturb} by $\alpha/X$ (see \cite{OConnor2010} for more details on the derivation of the hydrodynamic equations in \texttt{GR1D}.
\end{enumerate}
We note, however, that GR effects are not small within the PNS, where interior lepton number gradients can drive convection and affect the neutrino emission \citep{Wilson2003}. We leave a proper treatment of lepton number-driven convection to a future work.

The expression of $\omegaBV^2$, with the assumption of the background fluid being in hydrostatic equilibrium, is:

\begin{equation}
    \label{eq:BV_GR_noacc}
    \omegaBV^2 =  \frac{\alpha^2}{\rho h X^2} \frac{\text{d}\phi}{\dr}\left(\pder{\rho(1+\epsilon)}{r} - \frac{ 1}{ \rho c_s^2} \pder{P}{r} \right).
\end{equation}
For an accelerating fluid, the GR derivation becomes much less obvious, since non-inertial frames are involved. Fortunately, GR effects are very small in the outer layers, where the fluid has a non-negligible acceleration. Hence, we can simply use the Newtonian expression \eqref{eq:geff} with ad hoc corrections for relativistic space-time curvature, so that $\omegaBV^2$ becomes:

\begin{equation}
    \label{eq:BV_GR}
    \omegaBV^2 =  \frac{\alpha^2}{X^2} \left(\frac{\text{d}\phi}{\dr} - v \pder{v}{r}\right)\left(\frac{1}{\rho h} \pder{\rho(1+\epsilon)}{r} - \frac{ 1}{\rho h c_s^2} \pder{P}{r} \right),
\end{equation}
where now $v = X v_r$.

The main difference between Eqs.~\eqref{eq:BV_eff} and \eqref{eq:BV_GR} is the inclusion of $\partial \rho\epsilon / \partial r$ in the latter.  In the gain region the internal energy decreases with radius, i.e.~$\partial \rho\epsilon / \partial r < 0$.  This decreases the magnitude of $\omegaBV$ and therefore the amount of turbulence that is generated. We will come back to this in Section \ref{sec:resultsGR}
% {\color{red}  !!!!! Compare with simulations with effective potential and the internal energy included in omega before making strong claims !!!!!

% Let us now analyze in detail the differences in $\omegaBV$ with and without GR (i.e. Eqs. \eqref{eq:BV_GR} and \eqref{eq:BV_eff}). The GR coefficients $\alpha$, $X$ and $h = 1 + \epsilon + P/\rho$ are all close to $1$ in the post-shock region, where turbulence is the most significant. Therefore, eq. \eqref{eq:BV_GR} reduces almost exactly to eq. \eqref{eq:BV_eff}, the only (quite significant!) difference being the internal energy gradient added to the density gradient.

% As it turns out, despite $\epsilon$ being small compared to $\rho$, its gradient in the post-shock region is large enough to modify $\omegaBV$ significantly, and effectively making the star more (or sometimes less) stable against convection. }

\subsection{A word of caution: conservation of energy}
As briefly noted in CWO20, the equations of the \textit{STIR} model do not conserve energy. As pointed out by \cite{Muller2019_STIR}, the first and second term on the RHS of Eq. \eqref{eq:turb_kin_ene} break total energy conservation. Since turbulence is driven by neutrinos interacting with the material in the gain region, the rate of turbulent energy generation by buoyant driving is of the same order as the volume-integrated neutrino heating rate $\dot{\mathcal{Q}}_\nu$. This is also what \citep{Mabanta2019_turbulence_in_CCSN} assume in their turbulence model in the gain region, where the buoyant driving is given by $\beta \dot{\mathcal{Q}}_\nu$.

The extra energy coming from buoyant driving  breaks the energy-conserving structure of the hydrodynamical equations. As pointed out in  \cite{Mabanta2018_MLT_turb} this can be thought of as energy extracted from the free energy stored in unstable compositional and thermodynamic gradients. Therefore, although the energy is coming from a physically motivated phenomenon, it cannot  be consistently modeled in spherical symmetry, as recognized by \cite{Muller2019_STIR}.

If one looks at the problem of energy conservation in other artificially-driven explosions in one dimension, it is clear that energy is not conserved in any of them. This reflects the main issue with 1D models of CCSNe: they do not self-consistently explode. Hence, generally speaking, to drive an explosion in one dimension one needs more efficient neutrino heating. This extra-heating is injected in the model following different prescriptions, that can be based on physical phenomena \citep{Mabanta2018_MLT_turb,Couch2020_STIR} and/or observational constraints \citep{Ugliano2012,Sukhbold2016_explodability}, or simply by well-reasoned tweaking of some parameters \citep{OConnor2011_explodability,Perego2015_PUSH1}. Until multi-dimensional models become computationally more affordable, the above methods can give different and complementary insights into the physics behind the explosion mechanism of CCSNe.

\section{Results: Comparison with 3D simulations}
\label{sec:comparison3D}
\subsection{Results using an effective potential}
\label{sec:results_eff}
Following the approach of CWO20, we compare our GREP model to the \texttt{mesa20\_LR\_v} 3D simulation by \cite{OConnor2018_3Dprogenitors}. In the original \STIR/ paper, the advantage of carrying out such a comparison is that both \cite{OConnor2018_3Dprogenitors} and CWO20 use the code \texttt{FLASH}, so they can directly compare their 1D and 3D simulations with everything else unchanged. Despite using a different code, we still expect our GREP model to show very little differences to the original \STIR/, since \texttt{GR1D} and \texttt{FLASH} have been shown to yield similar results in 1D \citep{OConnor2018_comparison}.

In all the simulations presented in this section we used the same setup chosen by CWO20. That is, we simulated the collapse of a 20 M$_\odot$ progenitor from \cite{Farmer2016}, adopting the SFHo EOS \citep{Steiner2013_SFHo} and assuming Nuclear Statistical Equilibrium (NSE) everywhere. The neutrino radiation transport solver used in \texttt{FLASH} closely resembles the one used in \texttt{GR1D} \citep{OConnor2015}, and we used the same set of \texttt{NuLib} opacities adopted in CWO20. For the neutrinos, we used 12 energy groups geometrically spaced up to $\sim$250 MeV.

\begin{figure*}[hbt!]
    \centering
    \gridline{\fig{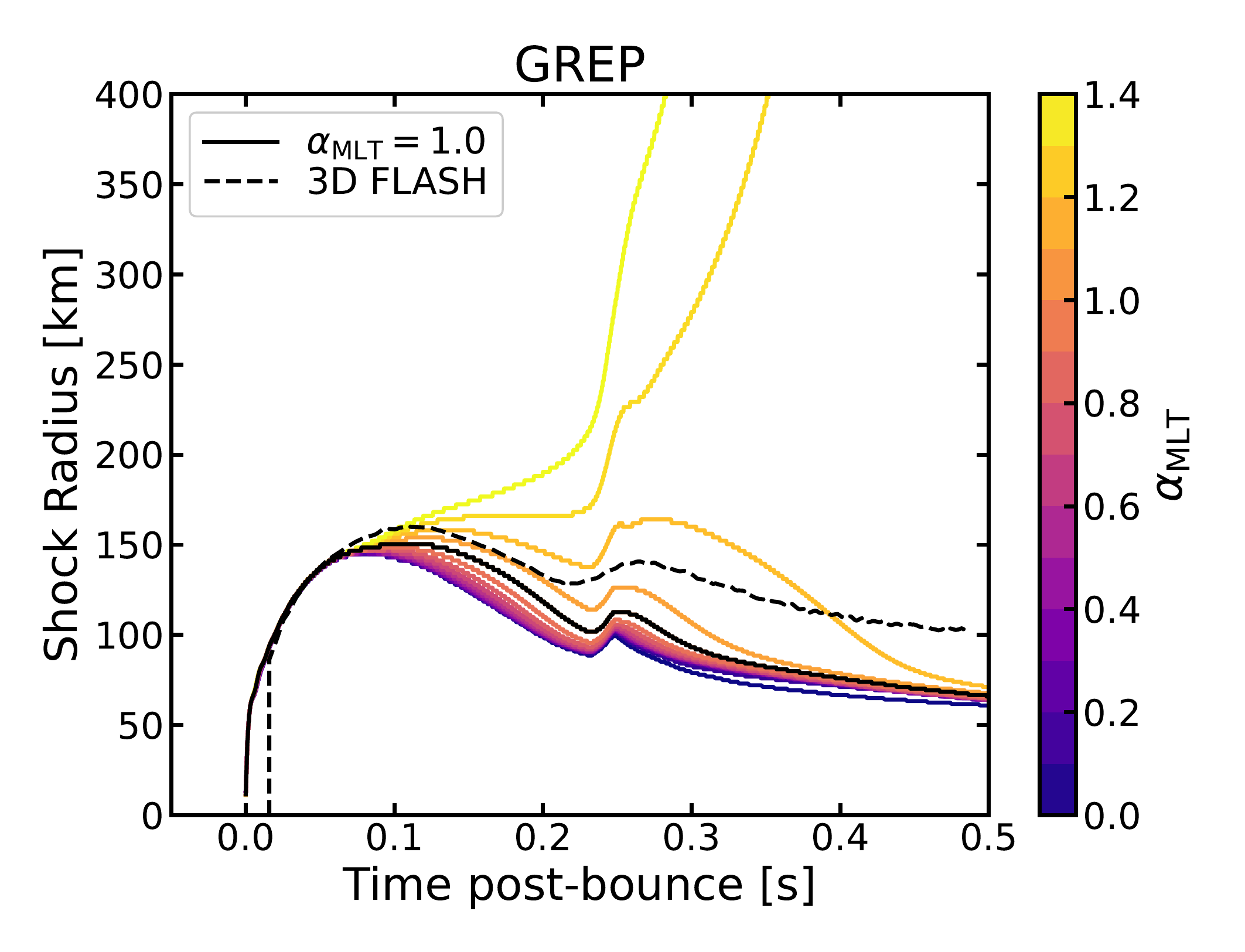}{0.5\textwidth}{(a)}
              \fig{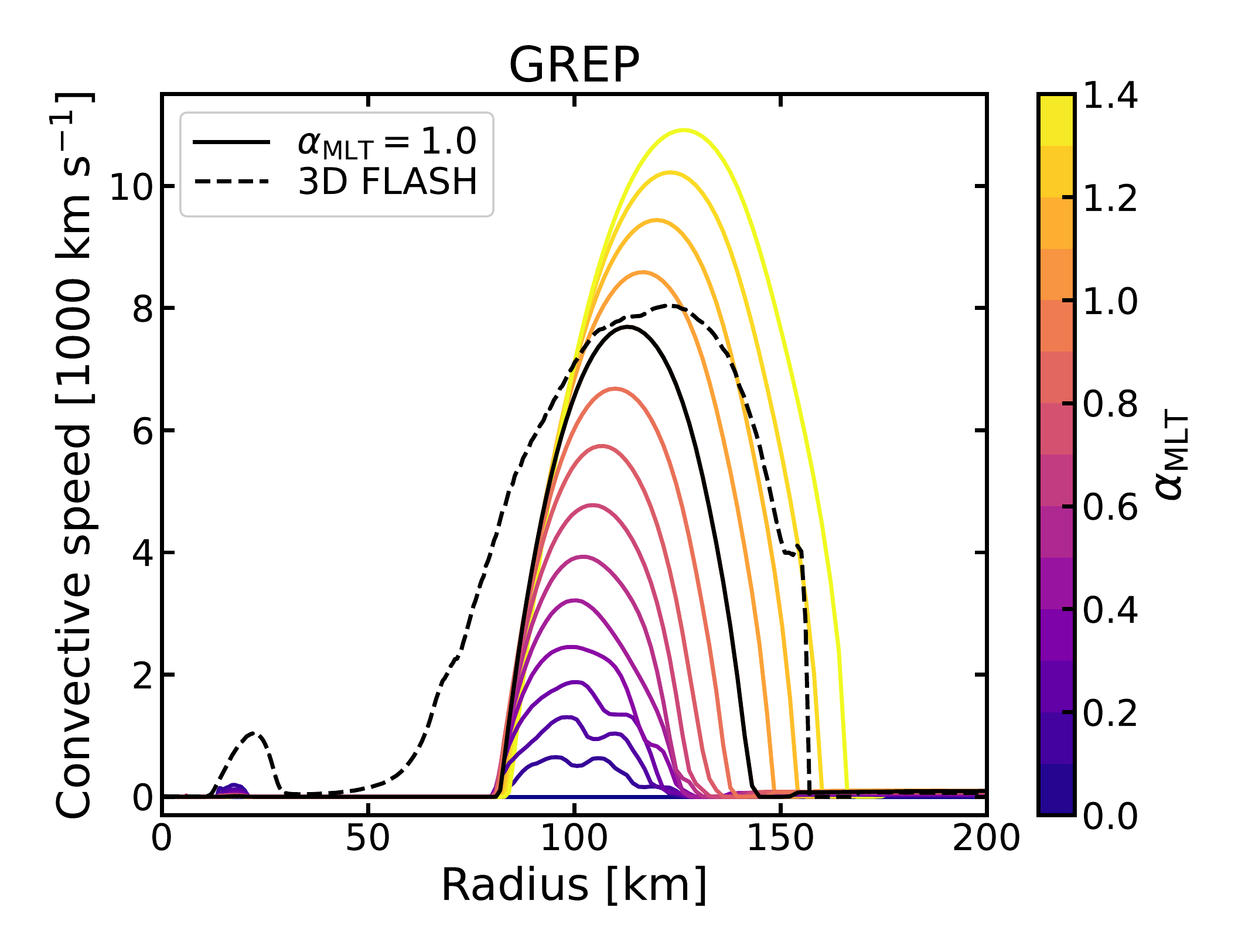}{0.5\textwidth}{(b)}}
    \gridline{\fig{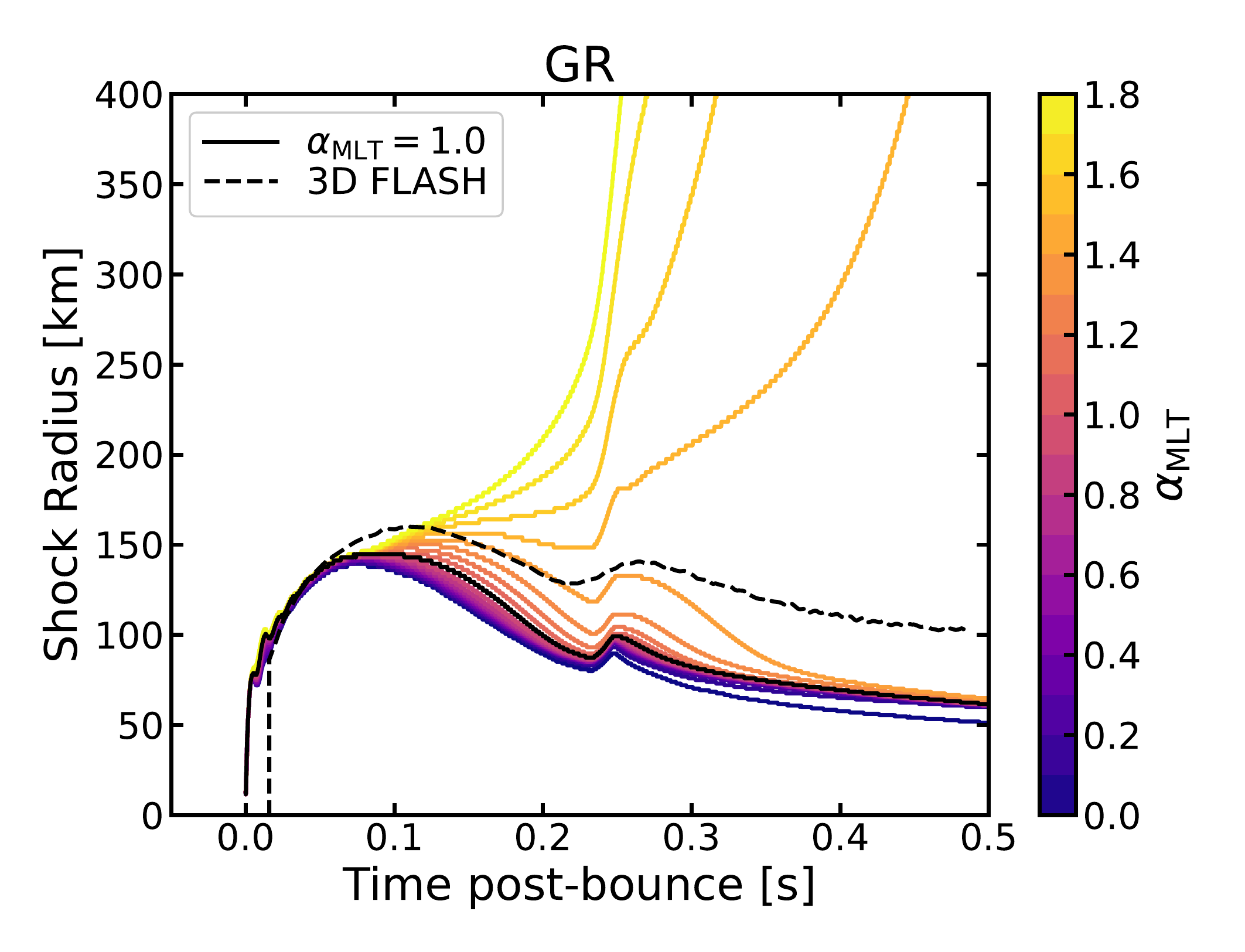}{0.5\textwidth}{(c)}
              \fig{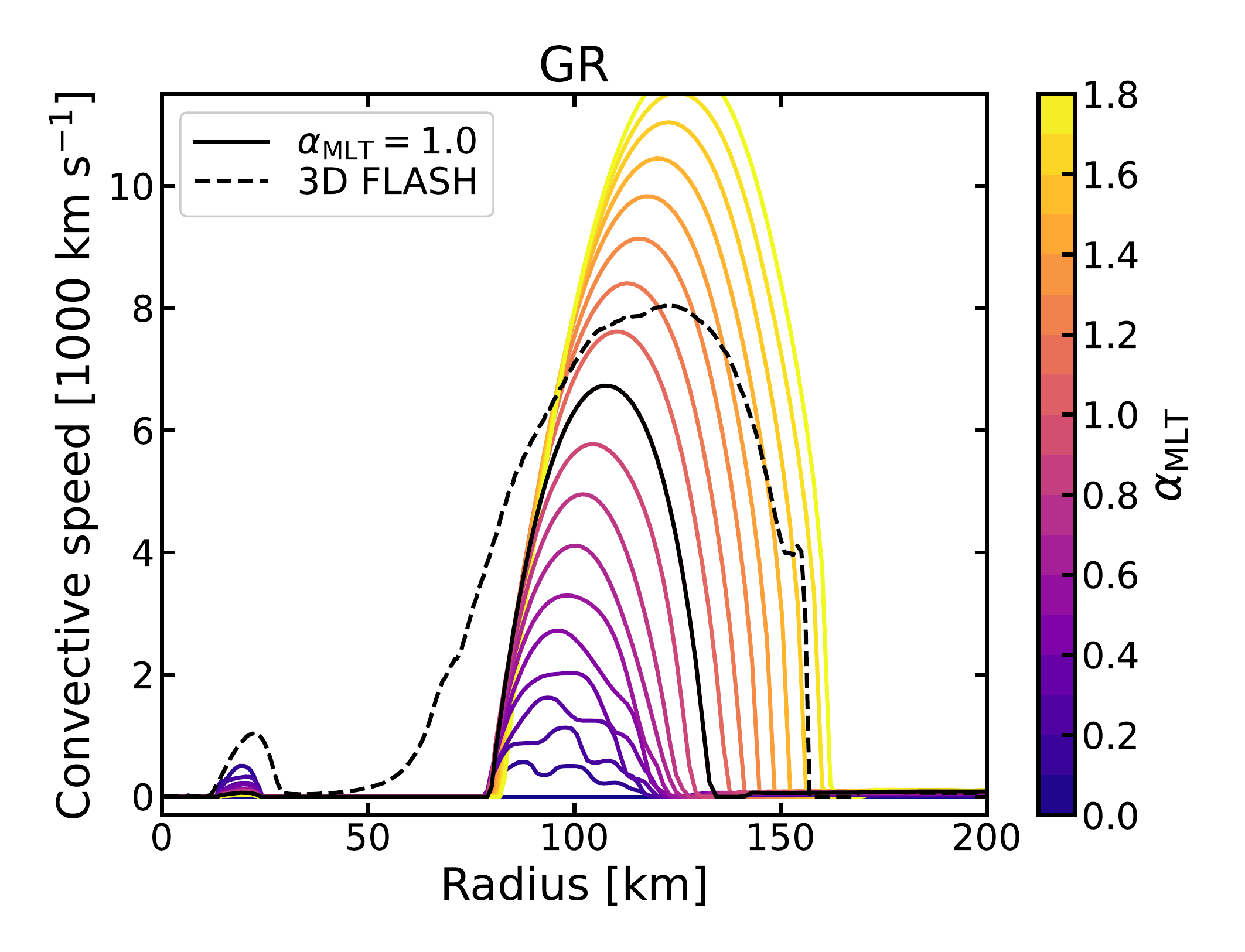}{0.5\textwidth}{(d)}}

    \caption{The plots on the upper row were generated using our GREP model, while the plots on the bottom row were generated using full GR. Panels (a,c) show the time evolution of the shock radius for different values of the parameter $\aMLT$, and can be compared to Figure 2 from \cite{Couch2020_STIR}. Panels (b,d) show a snapshot at $\sim 135$ ms post bounce of $\vturb$, and can be compared to Figure 1 from \cite{Couch2020_STIR}. The dashed lines are from the 3D simulation of \cite{OConnor2018_3Dprogenitors}.}
    \label{fig:turb}
\end{figure*}

In The upper panels of Figure \ref{fig:turb} we plot the shock radius versus time and the turbulent velocity profile at $\sim$135 ms post-bounce for our GREP model (to be compared to Figures 1 and 2 of CWO20). \texttt{GR1D} consistently yields larger values than \texttt{FLASH} for the turbulent velocity at a given $\aMLT$.  This translates into larger values of the shock radius at a given time. Except for these small differences, the two models agree very well, yielding explosions for $\aMLT > 1.2$. It is worth noting, however, that the 3D results show features that are not present in our MLT-like model. Specifically, the convective speed in 3D has a slower-decaying tail at ~50-80 km. As pointed out by CWO20, this is due to angular variations in the 3D model, rather than presence of convection in the region below the gain layer. As one would expect, in fact, convection in our model shuts off at ~80 km, which is approximately the location of the gain layer. More interesting is the lack of PNS convection at ~25 km, not captured by our model. A possible explanation for this is that we are not taking lepton number-driven convection into account, and with a more careful treatment of this type of convection, this feature might be correctly reproduced, but this goes beyond the scope of this paper.

\subsection{Results using GR}
\label{sec:resultsGR}
We approached the analysis of the simulations in full GR from two different points of view: on the one hand, we can compare the GR and GREP models with everything else being unchanged; on the other hand, we can compare the GR model to a 3D simulation in full GR. 

For the first part of the analysis, we show the results of our GREP model in the upper panels of Figure \ref{fig:turb}, while the bottom panels refer to the runs with full GR. The first and most important difference to notice is the range of $\aMLT$ used in the GREP and GR simulations. To achieve turbulent velocities and shock radii that are similar to the GREP results, the model in full GR requires an $\aMLT$ that is $\sim 20\%$ larger. To understand the origin of this increase in $\aMLT$, it's useful to look at the form of $\omegaBV$. As already pointed out in Section \ref{sec:STIR_GR}, the inclusion of the internal energy gradient in eq. \eqref{eq:BV_GR}, compared to eq. \eqref{eq:BV_eff}, is in fact the main difference between the GREP and GR models. 

In the gain region, which is where turbulence is most relevant, the internal energy gradient is negative, and therefore it decreases $\omegaBV^2$, making the fluid more stable against convection. The inclusion of $\rho \epsilon$ in the definition of $\omegaBV$ is therefore needed for a realistic characterization of turbulent convection. Indeed, if one takes the form of $\omegaBV$ from GR and implements it in the GREP model, the value of $\aMLT$ needed to achieve an explosion increases, becoming comparable to the one we use for the GR model. 
%However, the differences between our GR and GREP model cannot be simply explained by the different form of $\omegaBV$, as discussed in section \ref{sec:prog}.

The second part of the analysis was carried out by comparing the full GR model to the 3D octant simulations of \cite{Schneider2019}. For this part of the analysis, we simulated a 20 M$_\odot$ progenitor from \cite{WH07} with two of the EOS's used in \cite{Schneider2019}: SLy4$_{1.0}$ \citep{Chabanat1997_SLy4} and LS220 (a version of the original EOS with incompressibility $K_\infty = 220$ MeV from \cite{LS1991} recomputed using the SRO code from \cite{Schneider2017_SROEOS}). For the neutrino opacities, we adopted the same NuLib tables used by \cite{Schneider2019}, with 24 energy groups logarithmically spaced up to 280 MeV.

Like \cite{Schneider2019}, at 20 ms after bounce we switch from inelastic electron-neutrino scattering to a simpler elastic treatment, and we also shut off velocity-dependent terms in the transport equation. Unlike them, we consistently keep 24 energy groups throughout the whole simulation, while they switch to 16 energy groups at 20 ms post-bounce.  Our choice ensures a smooth transition in our code from inelastic to elastic electron-neutrino scattering.

\begin{figure*}[hbt!]
    \centering
    \gridline{\fig{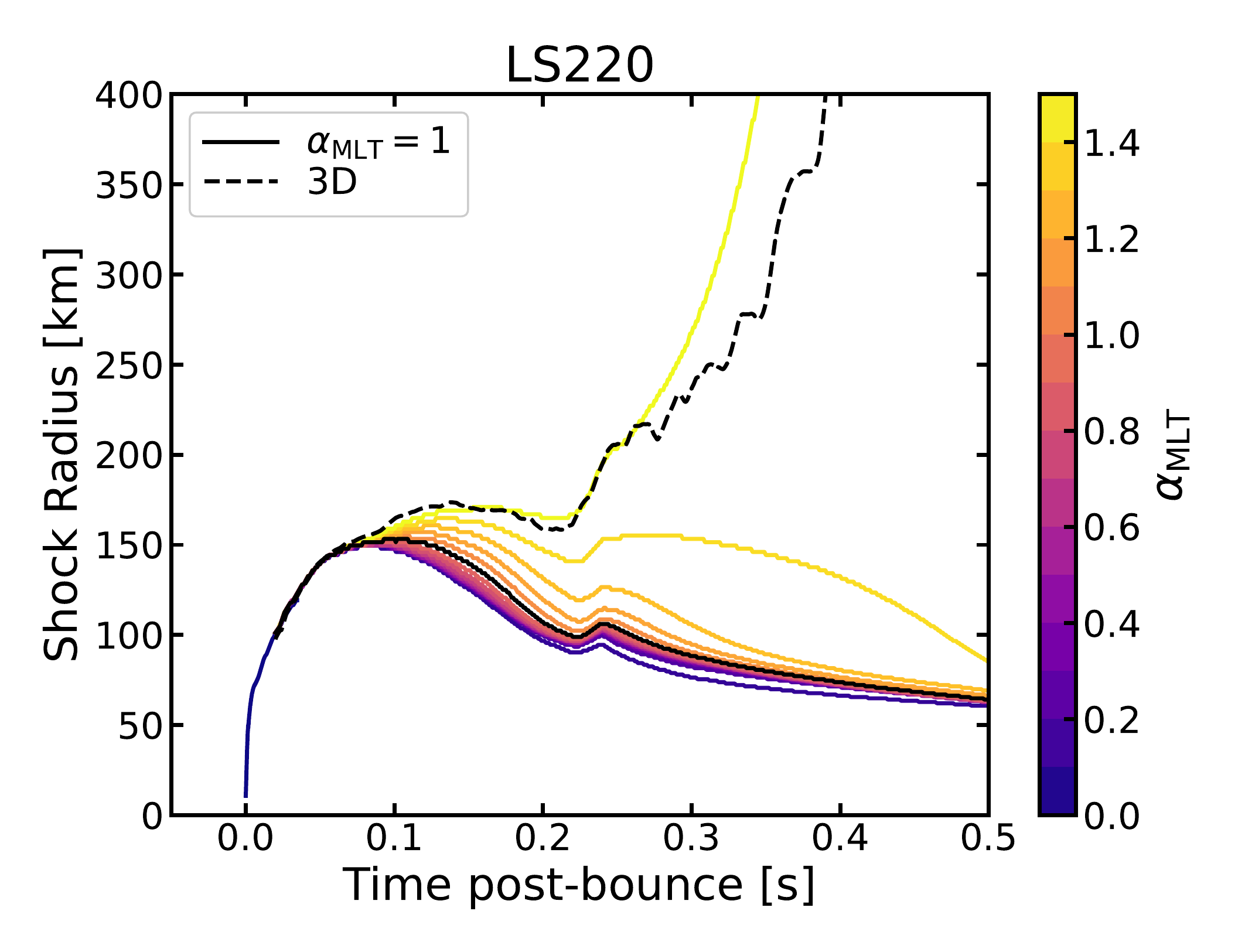}{0.5\textwidth}{(a)}
              \fig{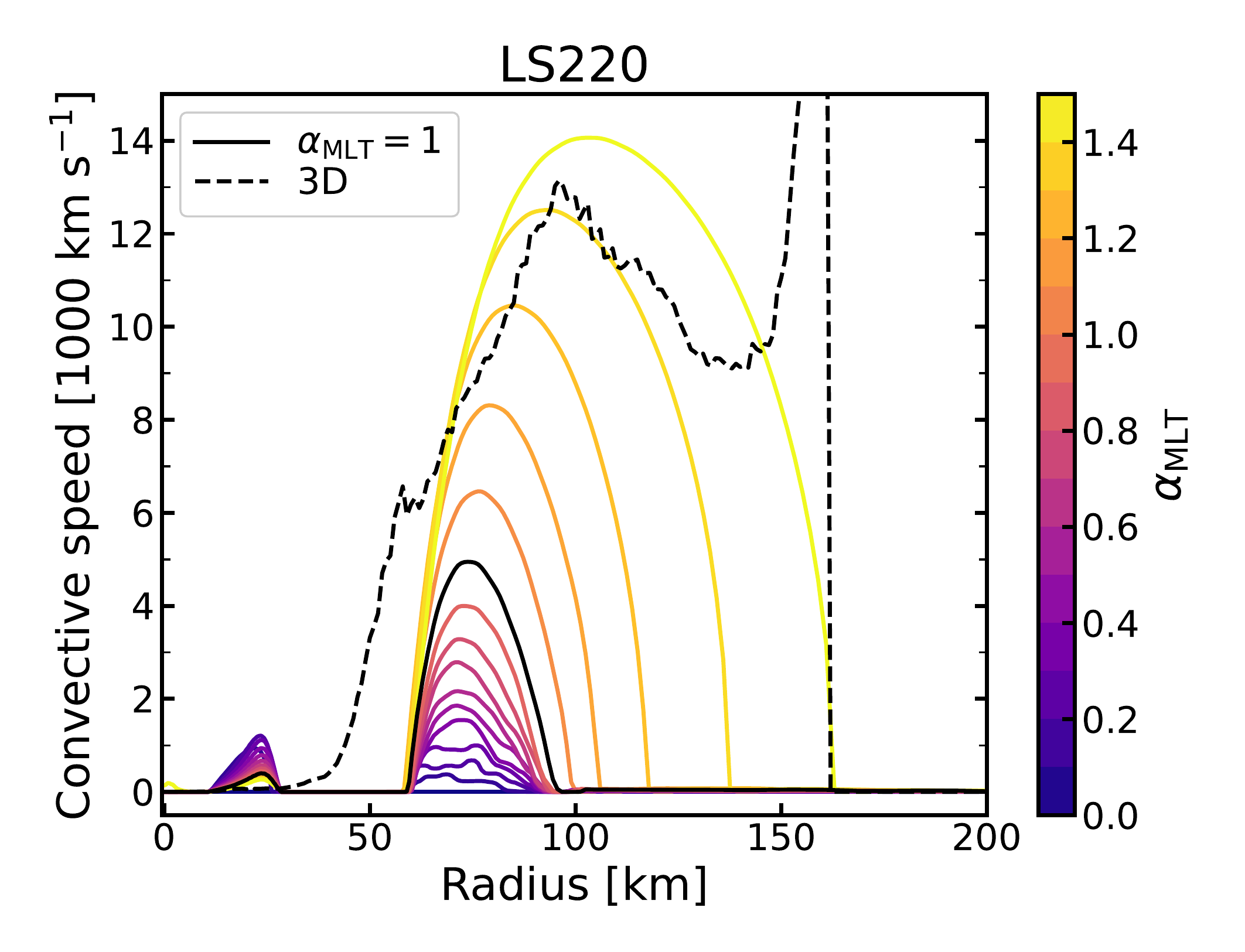}{0.5\textwidth}{(b)}}
    \gridline{\fig{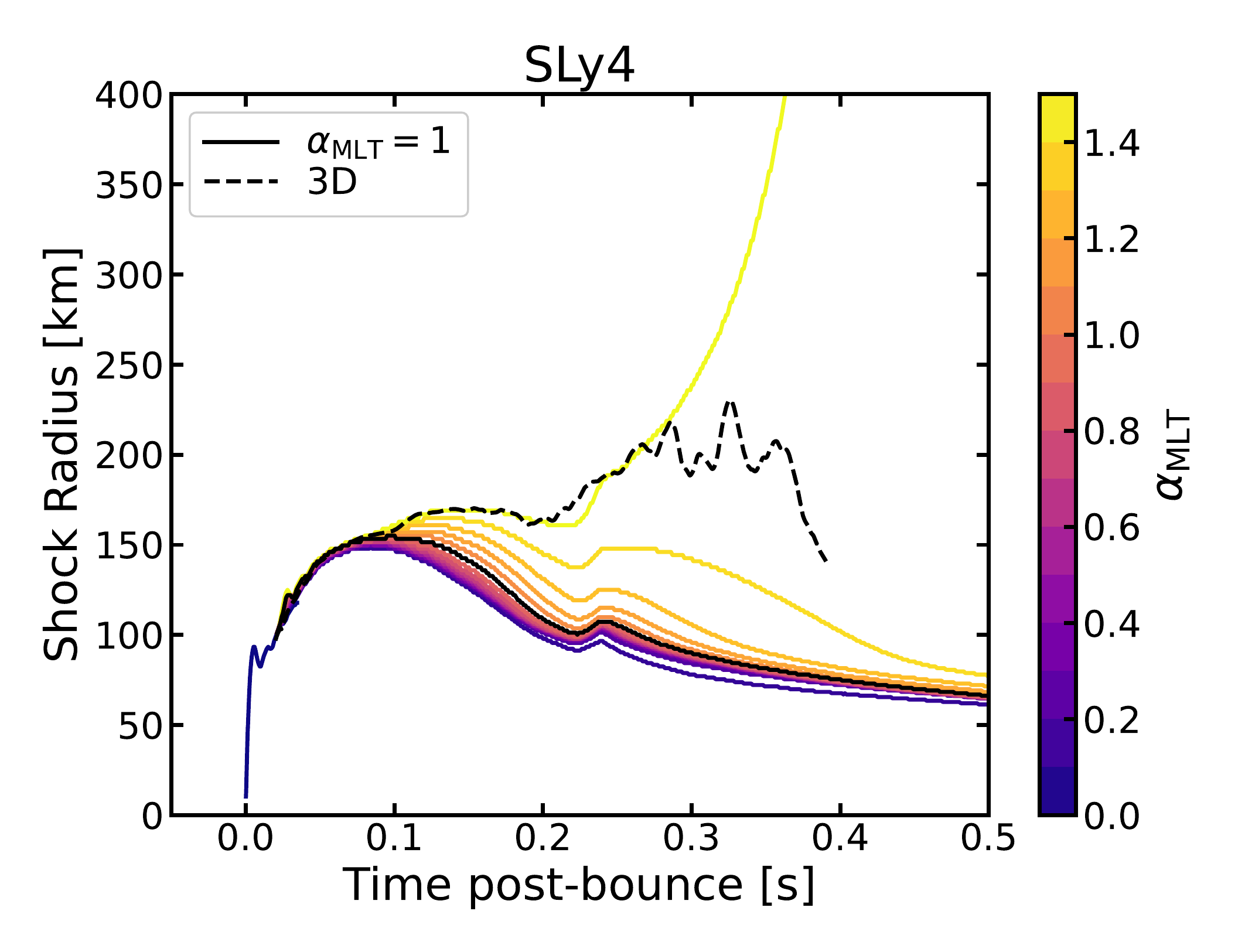}{0.5\textwidth}{(c)}
              \fig{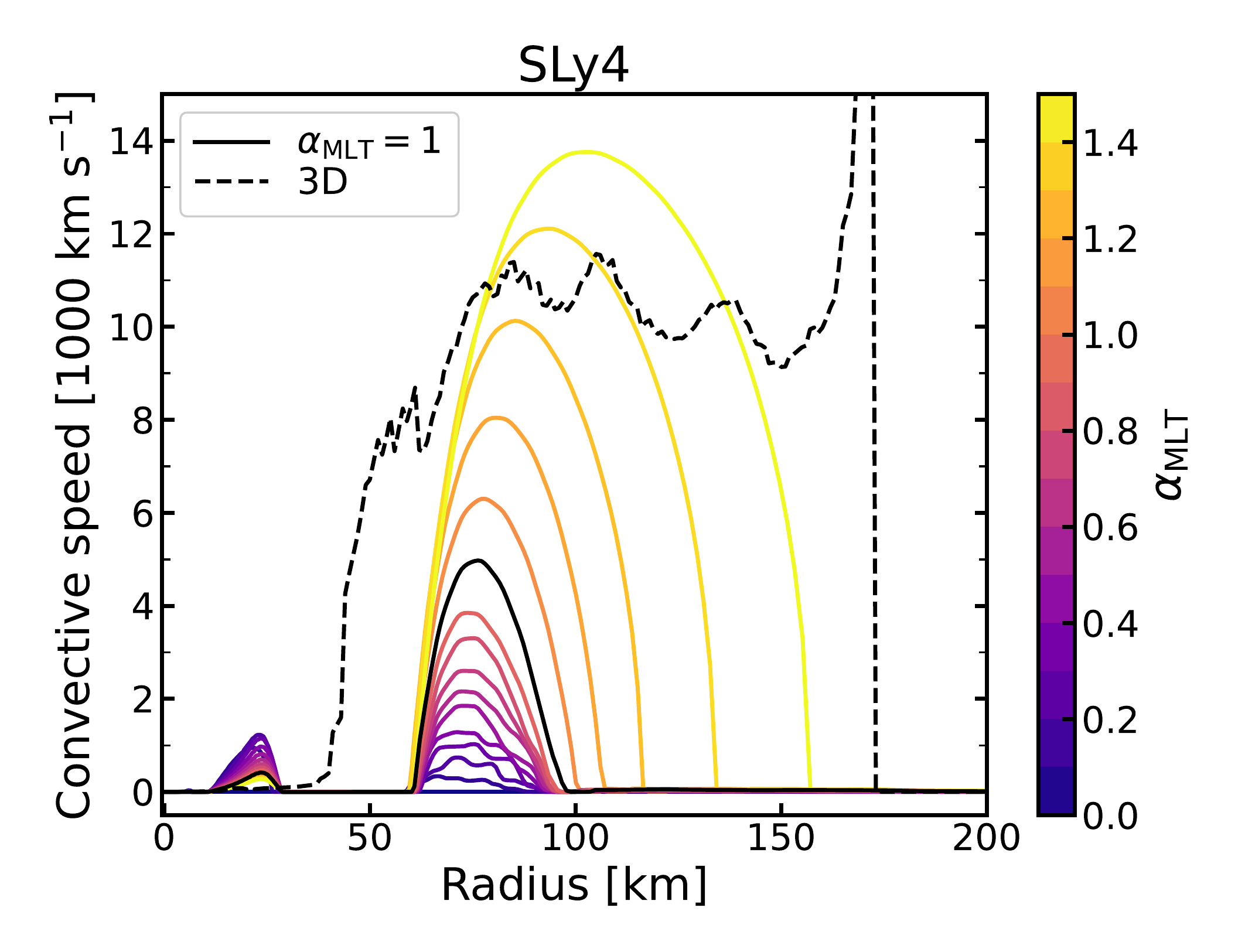}{0.5\textwidth}{(d)}}
              
    \caption{Comparison of our GR turbulent convection model (solid lines) as a function of $\aMLT$ with the 3D octant simulations from \cite{Schneider2019} (dashed lines). The upper (lower) panels refer to the simulations using the LS220 (SLy4) EOS. The panels on the left show the shock evolution, while the panels on the right show the turbulent velocity profile at $\sim 220$ ms after bounce.}
    \label{fig:compare3D_GR}
\end{figure*}

A comparison between the 3D data and our 1D model is presented in Figure \ref{fig:compare3D_GR}. As one can immediately see, for the LS220 case, both the shock radius and the turbulent velocity profile show very good agreement between the 3D and our 1D results. As already pointed out in section \ref{sec:results_eff}, minor differences in the turbulent velocity profile at small radii can be explained as the consequence of angular variations in the 3D simulation. The SLy4 case is a bit more complicated, and the agreements is not as good as in the LS220 case. After $\sim$300 ms the shock trajectory in 1D starts deviating from the 3D case, suggesting that turbulence in our MLT-like model might be slightly more efficient than in 3D, hence generating a successful explosion. A similar feature can be seen in the LS220 case, in which after $\sim$300ms the 3D shock trajectory drops below the one in 1D. Finally, it's worth mentioning that since the 3D simulations reported in Figure \ref{fig:compare3D_GR} were run imposing octant symmetry, SASI cannot develop. This represents a further confirmation that our GR model accurately predicts the development of turbulence in the pre-explosion phase of CCSNe.

%An interesting feature that can be seen in Figures \ref{fig:compare3D_GR_b} and \ref{fig:compare3D_GR_d} is the generation of turbulence just outside the PNS ($\sim20$ km), which is much more efficient in our GR model compared to the original \textit{STIR} formulation. In fact, by comparing Figure \ref{fig:turb_d} and Figure \ref{fig:turb_b}, one can see that for the \texttt{mesa20\_LR\_v} progenitor the convection above the PNS is much stronger in our GR model, while in the original \textit{STIR} there's much less turbulence generated at $\sim20$ km. {\color{red} Explain why in GR it's stronger, might be because you include epsilon and so the gradients right outside the PNS are stronger, but double check}. 

It is also worth noting that, for larger values of $\aMLT$, the turbulence inside the PNS is weaker.  This is consistent with the findings of CWO20. In their paper, they (semi-quantitatively) assert that the convection inside the PNS is too fast to allow a build-up of turbulence energy.  This is because the gradients are quickly flattened so there isn't enough time for turbulence to develop. This is the reason why for small values of $\aMLT$ our model is able to generate some turbulence outside the PNS, but as convection becomes more efficient (i.e. for larger values of $\aMLT$) the gradients are flattened and $\omegaBV^2$ becomes smaller, resulting in less turbulent energy generated above the PNS.

\subsection{Another word of caution: Resolution}
It is well known that, in 3D, low resolution simulations can overestimate the amount of turbulence produced in the gain region. The turbulence cascade is in fact artificially truncated at larger scales than what would otherwise happen in higher resolution simulations \citep{Radice2016}. Low resolution leads to an accumulation of turbulent energy at the largest scales, resulting in an additional pressure support in the gain region, which facilitates an explosion. In an MLT approach, this can typically be avoided, but other problems may arise (for a more detailed discussion see Appendix \ref{appendixRes}).

For the remainder of the paper, we chose to perform simulations at a higher spatial resolution and with more energy groups with respect to the results presented in Section \ref{sec:comparison3D}. As explained in Appendix \ref{appendixRes}, a higher resolution in space and in neutrino energy groups can change how much turbulent energy is injected in the model. However, it is quite unlikely that this will affect the \textit{qualitative} results  (i.e. explosion or fallback) of our simulation, because we are using a parametric model where one can choose how much turbulent energy will be injected in the model simply by changing the value of $\aMLT$. Therefore, if a simulation run with higher spatial resolution develops more (or less) turbulence, one can obtain the same qualitative result in a lower resolution simulation by increasing (or decreasing) the value of $\aMLT$. 

For the remainder of the paper, however, we chose to perform all of the simulations with a higher spatial resolution and with more neutrino energy groups with respect to Section \ref{sec:comparison3D}.

\section{Progenitor Study}
\label{sec:prog}
After validating our turbulent convection model by comparing it to the 3D results of \cite{Couch2014_3D} and \cite{Schneider2019}, we simulated the collapse and subsequent shock revival of 20 progenitors from \cite{Sukhbold2016_explodability} for different values of $\aMLT$, chosen such that the fraction of successful explosions is approximately between 25\% and 80\%. All of the simulations were run using a high spatial resolution, shown in the upper panel of Figure \ref{fig:resolution} as a dashed line, with 18 neutrino energy groups geometrically spaced from 0 to 280 MeV.

From the analysis carried out in section \ref{sec:comparison3D}, it is clear that once an explosion sets in for a certain value $\aMLT^*$, that progenitor will explode for all $\aMLT > \aMLT^*$ (see Figures \ref{fig:turb} and \ref{fig:compare3D_GR}). Therefore, if a given progenitor explodes for a given value of $\aMLT$, it will also explode for higher values of $\aMLT$. Analogously, if a given progenitor ends up in a failed explosion for a given value of $\aMLT$, it will also fail for lower values of $\aMLT$. Hence, we did not run simulations for those values of $\aMLT$ that could be already predicted by the outcome of the simulations at other values of $\aMLT$.

\begin{figure}[ht]
    \vspace{12pt}
    \centering
    \gridline{\fig{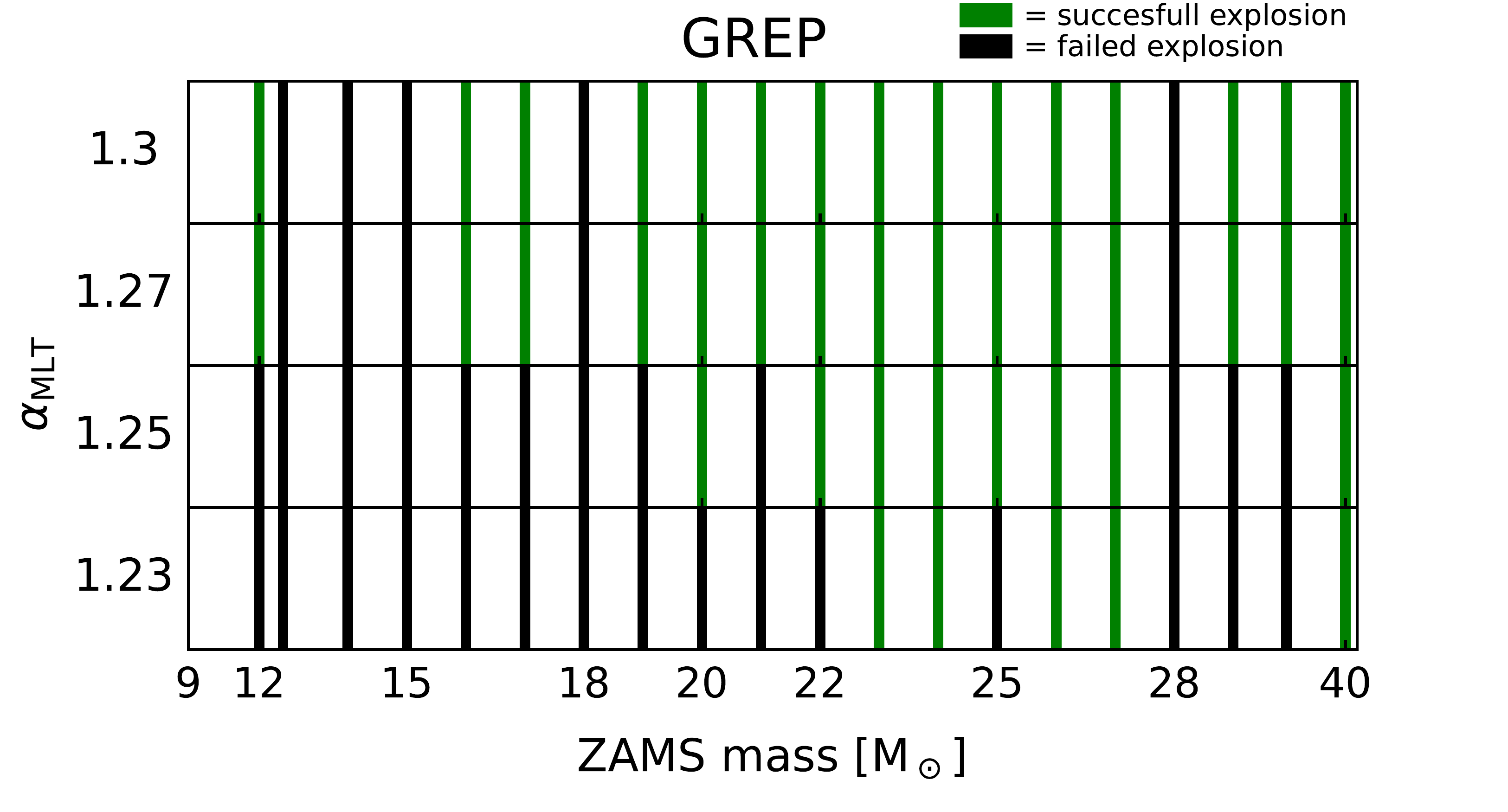}{\columnwidth}{(a)}}
    \gridline{\fig{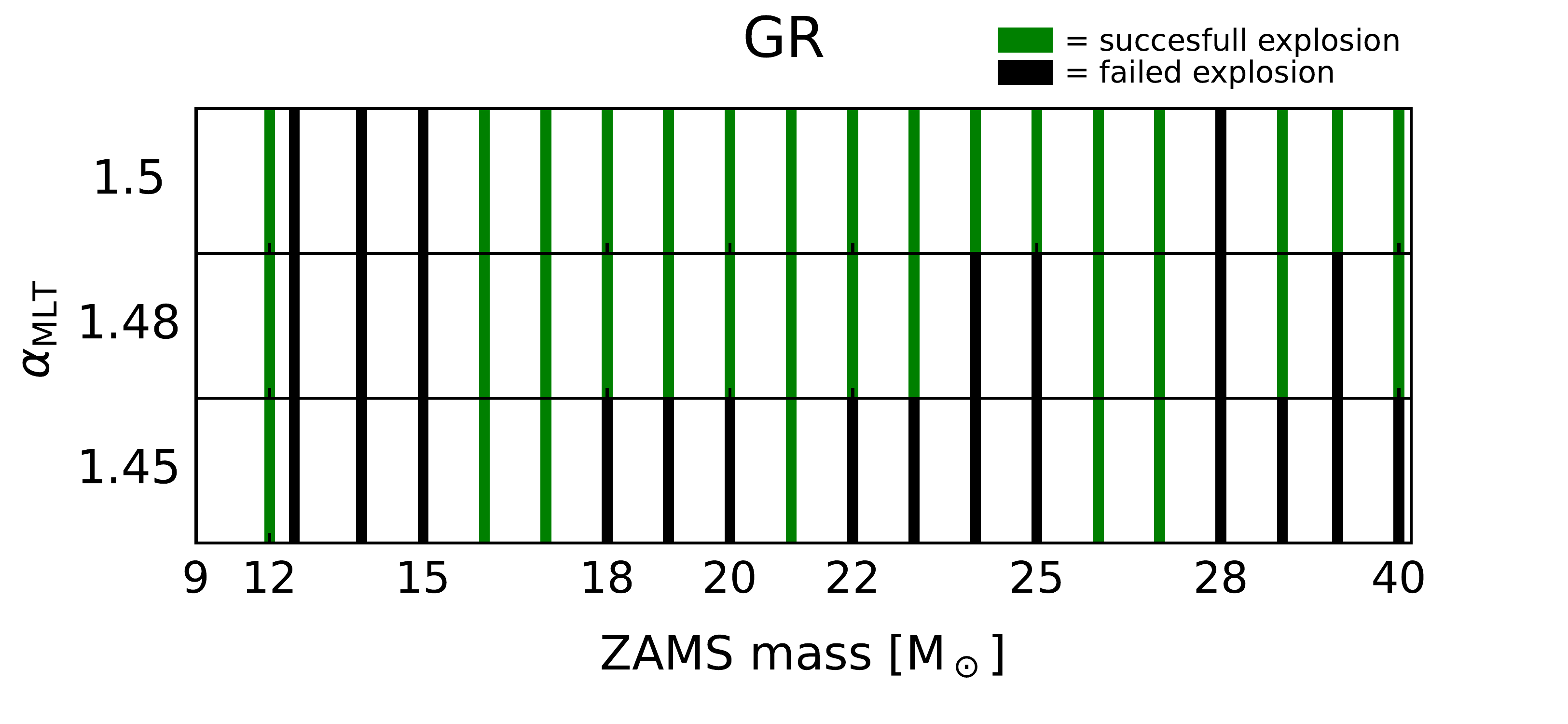}{\columnwidth}{(b)}}

    \caption{Explosion pattern of CCSN for the GREP (upper panel) and GR (lower panel) models as a function of the Zero Age Main Sequence mass. Green bands represent successful explosions (i.e. the shock has reached 500 km), while black bands represent failed explosions.}
    \label{fig:explodability}
\end{figure}

The results of our GREP model agree with the ones obtained by CWO20, although shifted by $\Delta \aMLT \simeq 0.05$, as can be seen by comparing the upper panel of Figure \ref{fig:explodability} with  Figure 6 from their paper. The shift is due to the fact that we use a  finer resolution in space and energy, and also to the fact that the numerical methods used to solve the hydrodynamics and the neutrino transport are slightly different.

More importantly, however, Figure \ref{fig:explodability} shows that the GR treatment of turbulent convection does not simply reproduce the \STIR/ results for a higher value of $\aMLT$, as Figure \ref{fig:turb} might suggest, but it modifies the explosion pattern of CCSNe. To accurately characterize the differences between the explodability patterns of the GR and GREP models, a study with more progenitors and a finer resolution in $\aMLT$ would be desired.  That, however,  is beyond the scope of the present work.

We can conclude from Figure \ref{fig:explodability}  that general relativity changes how likelihood for certain progenitors to explode. Focusing on the $\aMLT = 1.27$ and $\aMLT = 1.48$ panels, we find that the GREP model gives successful explosions for the 24, 25 and 30 M$_\odot$ progenitors, and failed explosions for the 18 M$_\odot$. In the GR model, the explodability is the exact opposite.
%Our GR results resemble more closely the explodability pattern obtained by \cite{Sukhbold2016_explodability} than the one obtained by CWO20.

%Since the major difference between the GR and GREP models is the form of the \BV/ frequency, we ran a few simulations with our GREP model adopting a ``GR-like" $\omegaBV$:
% \begin{equation}
%     \label{eq:BV_GR-like}
%     \omegaBV^2 =  g_\text{eff} \left(\frac{1}{\rho} \pder{\rho(1+\epsilon)}{r} - \frac{ 1}{ \rho c_s^2} \pder{P}{r} \right)
% \end{equation}
%For this model, one needs a larger value of $\aMLT$ to generates the same amount of turbulence as in the standard GREP, and this value is comparable to the one found in the GR model. As anticipated in section \ref{sec:resultsGR}, this is expected since the gradient of the internal energy in the gain region is negative, and therefore this reduces the magnitude of $\omegaBV$. 

The explodability pattern of the GR model with $\aMLT = 1.48$ is intermediate between the results of CWO20 and \cite{Sukhbold2016_explodability}: (i) like the former (and unlike the latter), we find that low mass progenitors with M = 13\,-15 M$_\odot$ result in failed explosions; (ii) like the latter (and unlike the former) we find that higher mass progenitors with M = 24\,-25 M$_\odot$ result in failed explosions. 

It is worth mentioning that the explodability pattern found in the GR model with $\aMLT = 1.48$ cannot be reproduced using the GREP model. Focusing on the upper panel of Figure \ref{fig:explodability}, one can notice how, in the $\aMLT = 1.25$ case, the 25 M$_\odot$ progenitor fails, but at the same time all  progenitors below 22 M$_\odot$ fail as well. This shows that the GREP model cannot produce successful explosions for low mass progenitors and at the same time failed explosions of the 24-25 M$_\odot$ progenitors. Notably, the GR model with $\aMLT = 1.5$ reproduces the same explodability pattern found in the GREP model with $\aMLT = 1.27$ (with the only exception of the 18 M$_\odot$ progenitor). This tells us that: (i) the threshold between successful and failed explosions is a very steep function of $\aMLT$; and (ii) GR can indeed reproduce the GREP results.  However, GR  also shows another possible pattern not yet observed in the GREP model of CWO20. 

Multi-D simulations cannot yet produce results analogous to Figure \ref{fig:explodability}, but the number of 3D simulations in the last 5 years has dramatically increased. In particular, a (relatively) large set of progenitors has been recently simulated by \cite{Burrows2020_3DFornax}. In their work, they simulate 14 progenitors from \cite{Sukhbold2016_explodability}, 12 of which are in the range $M=9-20$ M$_\odot$, and the remaining two have $M=25$ M$_\odot$ and $M=60$ M$_\odot$. Moreover, they use the same SFHo EOS that we employed for our progenitor study. However, like the vast majority of 3D simulations, they only include general relativistic effects in the hydrodynamics through the GREP model of \cite{Marek2006}, which they also use to account for redshift effects in the neutrino transport. Hence, we can qualitatively compare the outcome of our simulations with theirs. 

In \cite{Burrows2020_3DFornax} all of the progenitors explode except for the 13, 14 and 15 M$_\odot$, which is exactly what we find in the GREP model with $\aMLT = 1.27$. The GR model with $\aMLT = 1.48$ finds that the 25 M$_\odot$ should also fail, and it is interesting to note that in \cite{Burrows2020_3DFornax} that progenitor is the one with the  explosion occurring at the latest time, i.e. it is triggered at $\sim 400$ ms post bounce. Therefore, it can be considered a weak explosion, and a slightly less effective neutrino heating would cause the supernova to fail. This is consistent with what the GR model is showing.  That is,  among all  progenitors in the range 9-25 M$_\odot$, the 25 M$_\odot$ progenitor is the one that requires the largest value of $\aMLT$ to explode (i.e. $\sim 1.5$).
 
% {\color{red} Another type of analysis could be done by trying to quantify how "close to explosion" the progenitors are. Ertl (2016) have done several studies of that type, but I think it's (maybe) worth exploring in a different paper, where one could ask the question: "What does GR change during the explosion, that makes some progenitors closer to explosion w.r.t. GREP?"}

%However, this is not enough to explain the difference between the explodabilities of our GR and GREP models. In fact, if one looks at the explodability pattern generated by this ``GR-like" model, it's roughly the same as the one for the GREP model, as can be seen in Figure \ref{fig:GR-like_explodability}. Notice that Figure \ref{fig:GR-like_explodability} was generated using a lower resolution with respect to Figure \ref{fig:explodability}, but given our discussion in section \ref{sec:resolution}, this does not qualitatively impact our results. We can therefore conclude that simply modifying the \BV/ frequency is not enough to accurately reproduce the results of a model in full GR. 

This shows that including general relativistic effects can change how convective turbulence behaves in this  MLT model. We cannot predict if this effect translates to multi-dimensional simulations, but these results are suggesting that GR (rather than GREP) might have a significant impact on the explosion of CCSNe. Only a detailed comparison, done with 2D and 3D codes, between full-GR and Newtonian simulations, and across multiple progenitors, can clarify whether this effect translates to higher dimensions, where turbulent convection is generated self-consistently.

% As an example, one can look at Figure \ref{fig:compare3models}, where the 18 M$_\odot$ and 28 M$_\odot$ progenitors are compared for the GR, GREP and GR-like models. For both progenitors, the GR-like model is quite different from the simple GREP model, and it generates more turbulence than the other two models. However, for both progenitors, the GR-like model predicts the opposite outcome with respect to the GR model. Finally, it's clear that adjusting $\aMLT$ would only improve the agreement for one of the progenitors, while making the other worse.

\section{Conclusions}
In this article, we have shown how to extend \STIR/, the MLT-like model of CWO20, to a full general relativistic formalism. After verifying that we can reproduce the results of CWO20 using the same GREP model that they developed, we showed how a GR version of this model needs a larger value of $\aMLT$ to reproduce similar shock dynamics. The reason behind this is the inclusion of the internal energy gradient in eq. \eqref{eq:BV_GR}, which reduces the magnitude of $\omegaBV$, hence inhibiting the generation of turbulence, so that larger values of $\Lambda_{\rm mix}$ are needed to develop convective mixing as strong as in a Newtonian formalism.

The GR model reproduces very well the shock dynamics and turbulent energy generation in the octant 3D simulations of \cite{Schneider2019}, performed with two different equations of state.

After comparing our results to 3D results, we ran simulations of 20 different progenitors from \citep{Sukhbold2016_explodability} for different values of $\aMLT$, for both the GREP and GR models. We find that GR changes the explodability pattern of CCSNe. Specifically, the 24-25 M$_\odot$ progenitors seem to be hared to explode with the GR model. This results in an explodability as a function of progenitor mass that is intermediate  between the results of CWO20 and \citep{Sukhbold2016_explodability}. However, the GR model also shows, for $\aMLT = 1.5$, an explodability that is compatible with the results of CWO20.

\acknowledgments
The authors would like to thank Sean Couch, Andre da Silva Schneider and Mackenzie Warren for fruitful discussions.  Work at the University of Notre Dame supported by the U.S. Department of Energy under Nuclear Theory Grant DE-FG02-95-ER40934. EOC would like to acknowledge Vetenskapsr{\aa}det (the Swedish Research Council) for supporting this work under award numbers 2018-04575 and 2020-00452. 

%\vspace{5mm}
% \facilities{HST(STIS), Swift(XRT and UVOT), AAVSO, CTIO:1.3m,
% CTIO:1.5m,CXO}
% \software{astropy \citep{2013A&A...558A..33A},  
%           Cloudy \citep{2013RMxAA..49..137F}, 
%           SExtractor \citep{1996A&AS..117..393B}
%           }

\appendix 
\section{Derivation of the Brunt-V\"{a}is\"{a}l\"{a} frequency}
\label{appendixBV}
A derivation of $\omegaBV$ in general relativistic hydrostatic equilibrium can be found in Appendix C of \cite{Muller2013_3}. For clarity, here we go over a similar derivation, but for a different gauge, trying not to skip any non-obvious steps. In the fluid reference frame, the velocity is simply the convective velocity $v_c$. Hence, the 4-velocity is:

\begin{equation}
	u^\mu = (W/\alpha, Wv_c,0,0)
\end{equation}
Conservation of momentum requires that $\nabla_\mu T^{\mu r} = 0$, where:

\begin{align}
	\label{eq:Tmunu}
	T^{\mu\nu} &= (\rho h + \delta \rho h) u^\mu u^\nu + g^{\mu\nu} P \\
	T^{tt} &= \frac{(\rho h + \delta\rho h) W^2 - P}{\alpha^2}, \qquad T^{tr} = \frac{(\rho h + \delta \rho h) W^2 v_c}{\alpha X}, \qquad T^{rr} = \frac{(\rho h + \delta\rho h) W^2 v_c^2 + P}{X^2} \\
	T^{\theta\theta} &= \frac{P}{r^2}, \qquad T^{\phi\phi} = \frac{P}{r^2\sin^2\theta}
\end{align}
Note that the convective eddy is in pressure equilibrium with the surroundings. Therefore, there is a single pressure term in Eq. \eqref{eq:Tmunu}.
Writing out the conservation of momentum one gets:

\begin{equation}
	\label{eq:v_c_derivation1}
	\begin{split}
	\nabla_\mu T^{\mu r} &= 0 = \frac{1}{\sqrt{-g}}\partial_\mu (\sqrt{-g}T^{\mu r}) + \Gamma^r_{\rho\sigma} T^{\rho\sigma}\\
	&= \frac{1}{\sqrt{-g}}\partial_t (\sqrt{-g}T^{t r}) + \frac{1}{\sqrt{-g}}\partial_r (\sqrt{-g}T^{r r}) + \Gamma^r_{tt} T^{tt} + \Gamma^r_{rr} T^{rr} + \Gamma^r_{\theta\theta} T^{\theta\theta} + \Gamma^r_{\phi\phi} T^{\phi\phi} + 2 \Gamma^r_{rt} T^{rt} \\
	&= \frac{1}{\alpha X} \pder{}{t}\left[ (\rho h + \delta\rho h) W^2 v_c \right] + \frac{1}{\alpha X r^2} \pder{}{r}\left( \frac{\alpha r^2}{X}\left[ P + (\rho h + \delta\rho h) W^2 v_c^2 \right] \right) \\
	&+ \frac{1}{X^2}[(\rho h + \delta\rho h) W^2 - P]\partial_r \phi + \frac{P}{X^3}\partial_rX - \frac{r}{X^2} \frac{P}{r^2} - \frac{r \sin^2\theta}{X^2} \frac{P}{r^2 \sin^2\theta} + \frac{2(\rho h +\delta\rho h) W^2 v_c}{\alpha X^2} \partial_t X
	\end{split}
\end{equation}
The Christoffel symbols for the metric \eqref{eq:metric} can be found in Table A.1 from \cite{OConnor2010}.

By explicitly evaluating the derivative in the second term of Eq. \eqref{eq:v_c_derivation1}, it can be easily shown that:

\begin{equation}
	\label{eq:evaluate_dr_vc}
	\frac{1}{\alpha X r^2} \pder{}{r}\left( \frac{\alpha r^2}{X} P \right) = \frac{1}{X^2} \pder{P}{r} + \frac{P}{X^2} \pder{\phi}{r} + \frac{2P}{rX} - \frac{P}{X^3} \partial_rX
\end{equation}
In addition to this, using Eqs. A.5-7 from \cite{OConnor2010} one has:

\begin{equation}
	\label{eq:partialtX}
	v_c\partial_t X = v_c\frac{X^3}{r} \partial_t m \propto \cancelto{0}{v_c^2}
\end{equation}
By combining Eqs. \eqref{eq:v_c_derivation1} and \eqref{eq:evaluate_dr_vc}, and using $\rho h_t = \rho h + \delta \rho h$ we arrive at:

\begin{equation}
	\frac{1}{X^2}\pder{P}{r} + \frac{1}{\alpha X} \pder{}{t} [\rho h_t W^2 v_c] + \frac{1}{\alpha X r^2} \pder{}{r}\left( \frac{\alpha r^2}{X} \left[\rho h_t W^2 \cancelto{0}{v_c^2} \right]\right) = - \frac{\rho h_t W^2}{X^2}\pder{\phi}{r} ~~.
\end{equation}
Here we have ignored terms of order $v_c^2$. Now one can use the condition of hydrostatic equilibrium:

\begin{equation}
    \label{eq:hydrostatic_eq}
    \pder{P}{r} = - \rho h \pder{\phi}{r}.
\end{equation}
By considering subsonic motions in the gain region (hence $v_c << c_s << c \Rightarrow W^2 = 1$), one obtains:

\begin{equation}
	\label{eq:penultimatestep_BV}
	\frac{1}{\alpha X} \pder{}{t} \left(\rho h_t v_c\right) = - \frac{\delta\rho h}{X^2}\pder{\phi}{r}	= - \frac{1}{X^2}\pder{\phi}{r} C_L \delta r ~~,
\end{equation}
where for $C_L$ we have used the Schwarzschild discriminant from \cite{Thorne1966}:

\begin{equation}
	C_L = \left( \pder{\rho(1+\epsilon)}{r} - \frac{1}{c_s^2}\pder{P}{r} \right)~~.
\end{equation}
By using conservation of energy and rest mass (i.e. equations A.13 and A.17 from \cite{OConnor2010}), and assuming pressure equilibrium, it can be shown that:

\begin{equation}
    v_c \pder{\rho h_t}{t} \propto \cancelto{0}{v_c^2}    
\end{equation}
Therefore Eq. \eqref{eq:penultimatestep_BV} becomes:

\begin{equation}
	\pder{v_c^r}{t} = - \frac{\alpha}{\rho h_t X^2}\pder{\phi}{r} C_L \delta r ~~.
\end{equation}
Since $\ddot{r} = \alpha \dfrac{\partial v_c^r}{\partial t}$ (where the factor of $\alpha$ converts $t$ to proper time), we finally have:

\begin{equation}
	\ddot{r} = - \frac{\alpha^2}{X^2 \rho h_t}\pder{\phi}{r} C_L \delta r\; \Longrightarrow\; \omega_{\text{BV}}^2 = \frac{\alpha^2}{\rho h_t X^2}\pder{\phi}{r} \left( \pder{\rho(1+\epsilon)}{r} - \frac{1}{c_s^2}\pder{P}{r} \right)
\end{equation}

\begin{figure*}[ht]
    \centering
    \gridline{\fig{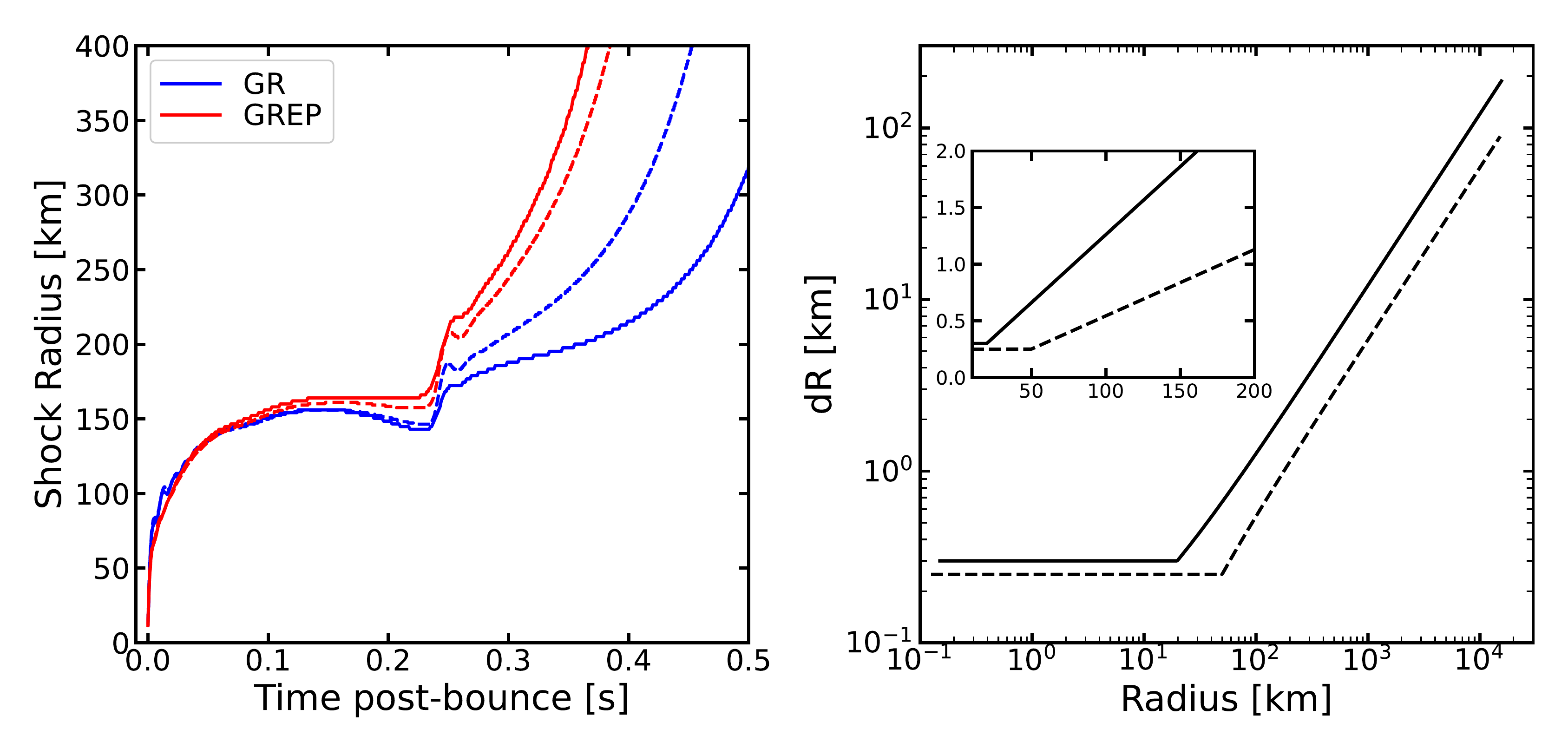}{\textwidth}{(a)}}
    \gridline{\fig{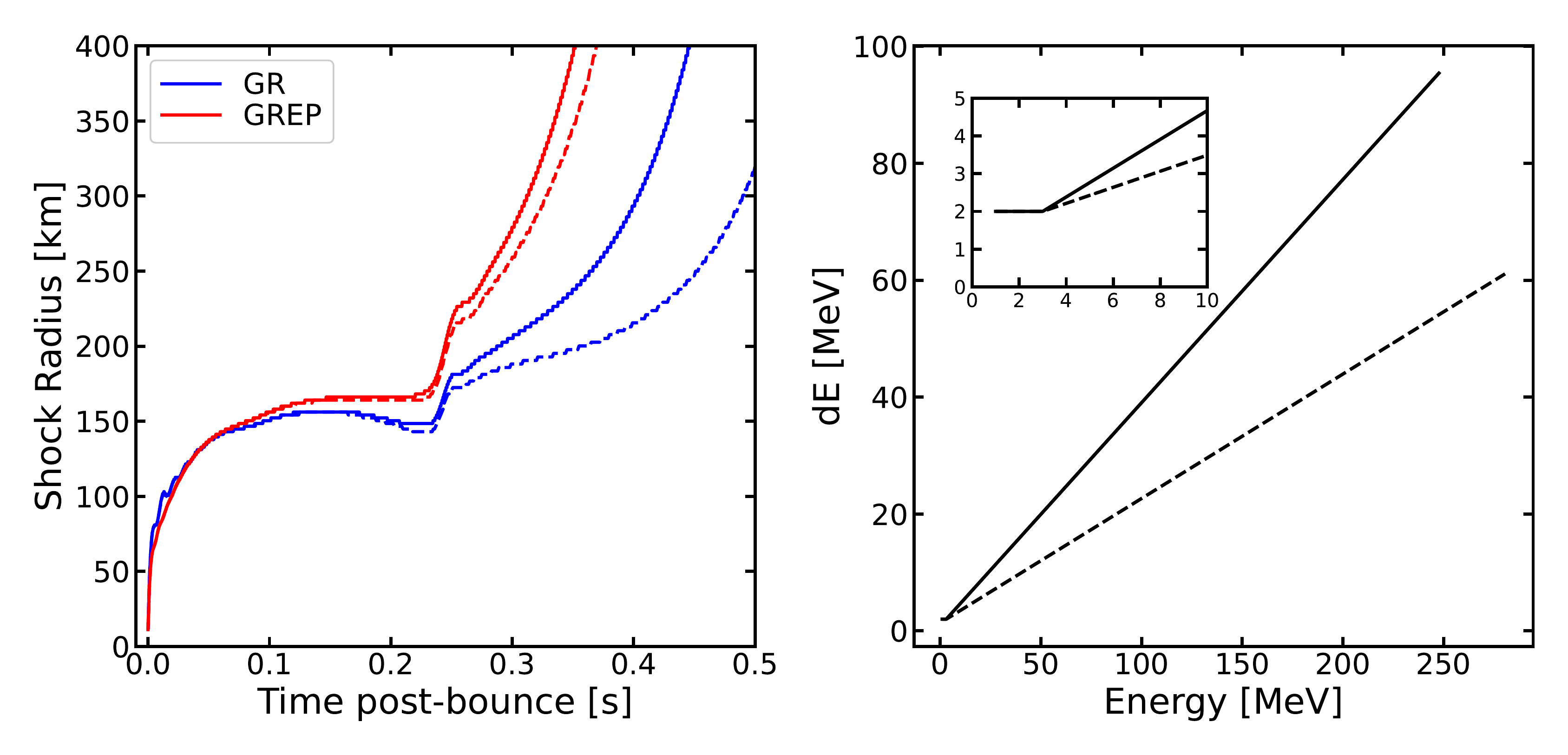}{\textwidth}{(b)}}
    
    \caption{Evolution of the shock radius for the \texttt{mesa20\_LR\_v} progenitor by \cite{OConnor2018_3Dprogenitors} for the GR (blue lines) and GREP models (red lines). The GR models were all run for $\aMLT = 1.5$, while the GREP models were all run with $\aMLT = 1.3$. (a) The upper panel compares low (solid lines) and high (dashed lines) spatial resolution simulations. All cases were run with 18 energy groups for neutrinos. (b) The lower panel compares simulations that use 18 (dashed lines) and 12 (solid lines) energy groups for neutrinos. These were run using the low resolution spatial grid shown as a solid line in panel (a)}
    \label{fig:resolution}
\end{figure*}

\section{Dependence of MLT-like turbulence on spatial and energy resolution}
\label{appendixRes}
Because turbulence is most effective behind the shock, it is crucial to have high resolution in the proximity of the shock, especially in the first few hundred milliseconds after bounce. In fact, in this time window, the shock is stalling at $\sim150$ km, and turbulent energy can provide the extra-pressure necessary to trigger the explosion. In our numerical treatment of equation \eqref{eq:vturb} we use a 5-point interpolation formula to evaluate the gradient of $v_r$, and to avoid numerical instabilities we shut off turbulence within two zones from the shock. This ensures that the turbulence is not advected through the shock, which, being a discontinuity in pressure and radial velocity, would make the evaluation of gradients problematic. However, this means that in our model there is no turbulence in the two zones right behind the shock. Therefore, if the resolution behind the shock is poor (typically $\Delta r \sim 1/1.5$ km at $r\sim 100/150$ km), one ends up shutting off turbulence in the most important regions (i.e. close to the shock).

We ran some tests by changing the spatial resolution for the GREP and GR model (see upper panel of Fig. \ref{fig:resolution}) for the same \texttt{mesa20\_LR\_v} progenitor used in Section \ref{sec:comparison3D}. From these tests it appears that, with the increased resolution in the GR model, the shock is pushed to larger radii at early times. This allows a widening of the gain region, generating a stronger explosion. This feature isn't present in the GREP model, which on the contrary shows weaker explosions at higher spatial resolutions. We also ran some tests by changing the resolution in energy.  For this case the GR and GREP models showed the same behaviour, i.e. higher energy resolution yields weaker explosions (see lower panel of Fig. \ref{fig:resolution}).

The dependence of the shock dynamics on resolution is not as straightforward as what might be expected based on simple arguments. However, the effects are relatively small, limited to a few percents in $\aMLT$, and therefore do not significantly affect our analysis.

\bibliographystyle{aasjournal}
\bibliography{References_SN,References_CNO,References_Nucleosynthesis,References_EoS_nuRates}

\end{document}